\documentclass[aps, showpacs, twocolumn]{revtex4}
\usepackage{graphicx}
\usepackage{dcolumn}
\usepackage{amsmath}
\usepackage{amssymb}
\usepackage{ulem}
\usepackage[colorlinks]{hyperref}
\hypersetup{citecolor=blue}
\usepackage{bm}
\usepackage{epstopdf}
\usepackage{amsthm}
\usepackage{float}

\usepackage[caption=false]{subfig}

\usepackage{mathrsfs}
\usepackage{multirow, array, booktabs}

\def\rmp#1#2#3{{ Rev. Mod. Phys.} {\bf #1}, #2 (#3)}
\def\prl#1#2#3{{ Phys. Rev. Lett.} {\bf #1}, #2 (#3)}
\def\pra#1#2#3{Phys. Rev. A {\bf #1}, #2 (#3)}
\def\prb#1#2#3{Phys. Rev. B {\bf #1}, #2 (#3)}

\def\pre#1#2#3{Phys. Rev. E {\bf #1}, #2 (#3)}

\def\epl#1#2#3{{ Euro. Phys. Lett.} {\bf #1}, #2 (#3)}
\def\epjb#1#2#3{{ Euro. Phys. J. B} {\bf #1}, #2 (#3)}
\def\prep#1#2#3{{ Phys. Rep.} {\bf #1}, #2 (#3)}

\def\physa#1#2#3{Physica A {\bf #1}, #2 (#3)}

\def\natcom#1#2#3{Nat. Commun. {\bf #1}, #2 (#3)}
\def\natphys#1#2#3{Nat. Phys. {\bf #1}, #2 (#3)}
\def\sc#1#2#3{Science {\bf #1}, #2 (#3)}

\def\jphya#1#2#3{J. Phys. A {\bf #1}, #2, (#3)}

\def\etl{$et~al.~$}
\def\etc{etc.}
\def\la{\langle}
\def\ra{\rangle}

\usepackage{xcolor}
\definecolor{C0}{RGB}{31,119,180}
\definecolor{C1}{RGB}{255,127,14}
\definecolor{C2}{RGB}{44,160,44}
\definecolor{C3}{RGB}{214,39,40}
\definecolor{C4}{RGB}{148,103,189}

\def\beqr{\begin{eqnarray}}
\def\eqnr{\end{eqnarray}}
\def\beq{\begin{equation}}
\def\bc{\begin{center}}
\def\ec{\end{center}}
\def\eqn{\end{equation}\noindent}
\begin{document}

\title{Extreme events scaling in self-organized critical models}

\author{Abdul Quadir}
\email{abdulq2013@gmail.com}
\author{Haider Hasan Jafri}
\email{haiderjaf@gmail.com}
\affiliation{Department of Physics, Aligarh Muslim University, Aligarh, 202 002, India}

\begin{abstract}
We study extreme events of avalanche activities in finite-size two-dimensional self-organized critical (SOC) models, specifically the stochastic Manna model (SMM) and the Bak-Tang-Weisenfeld (BTW) sandpile model. Employing the approach of block maxima, the study numerically reveals that the distributions for extreme avalanche size and area follow the generalized extreme value (GEV) distribution. The extreme avalanche size follows the Gumbel distribution with shape parameter $\xi=0$ while in the case of the extreme avalanche area, we report $\xi>0$. We propose scaling functions for extreme avalanche activities that connect the activities on different length scales. With the help of data collapse, we estimate the precise values of these critical exponents. The scaling functions provide an understanding of the intricate dynamics for different variants of the sandpile model, shedding light on the relationship between system size and extreme event characteristics. Our findings give insight into the extreme behavior of SOC models and offer a framework to understand the statistical properties of extreme events.

\end{abstract}

\maketitle

\section{Introduction}

A complex system refers to a system composed of numerous interconnected components or elements. The interactions among these components give rise to emergent behavior that is often difficult to predict from the properties of the individual parts. The widespread existence of complex networks in both natural and societal contexts, such as interlinked biological and chemical systems, neural networks, the internet, the WWW, and social networks, indicates that complexity is pervasive~\cite{Dorogovtsev_2003, Caldarelli_2007, Barrat_2008, Barabasi_2016, Albert_2002}. Because of the fluctuating nature of these systems, with their inherent intricacies and differences in properties, they have been a subject of continuing interest in the past few decades. The dynamics of complex systems may give rise to large fluctuations in a relevant variable, resulting in extreme events that may be defined as events exceeding a predefined large threshold. These events may have cascading effects throughout the system and may deviate significantly from a system's average (or usual) patterns. They are often rare but can have profound and disproportionate impacts. Extreme events occur in numerous scenarios, namely, the breakdown of a mechanical structure, an earthquake, flooding or crashes in financial markets \cite{Montroll_1974}. Other instances where these events have been reported are the systems that exhibit self-organized critical phenomena~\cite{BTW, Tang_1988, Bak_1996, Dhar_1989, Dhar_1990, Dhar_2006}, spontaneous brain activity, fracture~\cite{Sahini_1993}, portfolio management, Darwinian evolution of fitter proteins~\cite{Weinreich_2006}, fickle stock exchange~\cite{Goncalvesab_2011}, acute scenarios in capricious weather~\cite{Cai_2014, Katz_1992, Yao_2022}, seismicity risk evolution and other geophysical processes.

In the complex systems theory framework, it is important to understand and classify the possible underlying mechanisms responsible for huge fluctuations. Further, one can also derive statistical properties of the underlying statistical distributions. The classical theory that offers a systematic approach to understand the statistical properties of rare (or extreme) events is called as the extreme value theory (EVT). This allows for a deeper exploration of tail behavior beyond the scope of conventional statistical methods. EVT has been studied in engineering~\cite{Gumbel_1958, Weibull_1951}, finance~\cite{Embrecht_1997}, hydrology~\cite{Katz_2002}, meteorology~\cite{Storch_2002}, pinning-depinning dynamics~\cite{Yan_2024} and natural sciences~\cite{Katz_2002, Storch_2002, Yan_2024, Gutenberg_1944, Bouchaud_1997}, to name a few. Traditionally, the EVT is categorized into three extreme value statistics limit distributions: Fr\'echet, Fisher-Tippett-Gumbel (FTG), and  Weibull. In the literature, numerous studies have focused on the study of finite size effects using the renormalization-group (RG) theory and proved that RG applies regardless of whether these variables are independent or correlated~\cite{Coles_2001, Haan_2006, Galambos_1978, Albeverio_2006, Calvo_2012, Fortin_2015, Bramwell_2009, Castillo_2005, Clusel_2008, Bramwell_2000, Bertin_2005, Ghil_2011, Schehr_2006}. In $2-$dimensional site percolation problem, since the cluster size in sub-critical site percolation follows $p(x) \sim x^{-1}e^{x/x_c}$, with $x_c$ as cutoff~\cite{Stauffer_1994}, the largest cluster size follows the FTG distribution~\cite{Gyorgyi_2008, Gyorgyi_2010}. The EVT in $1/f^{\alpha}$ suggests FTG distribution for $0 \leq \alpha < 1$ whereas it follows a nontrivial distribution for $\alpha>1$~\cite{Antal_2001, Antal_2009}. In the case of the fitness model for the scale-free networks having nodes with homogeneous fitness, the degree distribution converges to the Gumbel distribution. However, the distribution converges to the Fr\'echet distribution in the case of nodes having heterogeneous fitness~\cite{Moreira_2002}. 

The presence of many degrees of freedom in the complex system leads to the critical phenomenon and the emergence of long-range correlation. Such long-ranged phenomena that exhibit scale invariance in the absence of an external tuning parameter are known as  ``Self-organized criticality (SOC)''. The SOC systems are characterized by their tendency to evolve towards a critical state~\cite{BTW}. In this situation, the observable quantities display power-law distributions. P. Bak, C. Tang, and K. Wiesenfeld's (BTW) hypothesis elucidates the emergence of scaling in a slowly driven non-equilibrium system found in nature~\cite{BTW, Bak_1996, Tang_1988}.  The SOC systems are an important class of complex systems that are capable of generating extreme events. Understanding extreme events in the SOC systems is important because it provides insight into the robustness and vulnerability of these systems, helps assess the potential risks associated with rare events, and further contributes to the understanding of self-organizing systems.

Additionally, the study of extreme events in the SOC systems has implications for risk management, disaster preparedness, and the resilience of systems in the face of unpredictable and impactful events. The Abelian BTW sandpile with infinite system size is critical, and the behavior of extreme activities can be easily derived~\cite{Katz_1992}. However, in real situations, the systems have a finite size, which may induce correlations, cutoffs and system size effects in the events~\cite{Dhar_1989, Dhar_2006, Dhar_1990, Yadav_2022}. Moreover, moment analysis of avalanche distributions has revealed multiscaling behavior in certain SOC models.
This is particularly evident in deterministic models, such as the BTW sandpile~\cite{Chessa_1999}. This finding indicates that scaling in such systems can be more complex and model-dependent than simple power-law behavior implies.

In this work, we show that the extreme value distribution (EVD), associated with extreme activity, scales with the system size. We highlight that the probability and parameters may vary with the system size and belong to the same class of generalized extreme value distributions (GEVD).  We demonstrate a simple scaling or data collapse function that can capture this characteristic. The method depends on identifying the characteristics of scaling functions for GEVD.  The rest of the article is organized as follows. In Sec.~\ref{Sec:GEVT}, we describe the generalized extreme value distribution. In Sec.~\ref{sec-model}, we study the SMM and the Abelian BTW sandpile model to study the extreme avalanche activities. Further, we propose the finite-size (FS) scaling for extreme avalanche activities in Sec.~\ref{Sec:USF}. Finally, in Sec.~\ref{Sec:Conclusion}, we draw our conclusions.

\section{Generalized Extreme Value Theory}~\label{Sec:GEVT}
To understand the statistics of extreme events, we consider the distribution of the maxima of the observable quantity. The maxima are obtained by dividing the dataset into intervals of fixed length. Then the maximum value from each block is considered. Let $x_i \in max\{X_1, X_2,... \}$ be the maxima of the independent and identically distributed random observables in each block $X_i$. It was shown that the distributions of maxima $X_i$ may follow a single peak probability distribution~\cite{Gumbel_1958}. The cumulative distribution function (CDF) for the maxima $x_i$ results in the GEV distribution, which is given as~\cite{Coles_2001, Galambos_1978, Haan_2006, Albeverio_2006}
\beq
\mathcal{F}(x,\mu,\beta,\xi) = \exp \Big\{ - \Big[ 1 + \xi \Big( \dfrac{x - \mu}{\beta} \Big) \Big]^{-1/\xi} \Big\} ~\label{Eq:GEVD1}
\eqn
where $\mu,~\beta$ and $\xi$ are location (or mode), scale, and shape parameters, respectively, having bounds $ \mu,~\xi \in \mathbb{R} $ and $ \beta \in \mathbb{R} \mid \beta > 0 $. Depending upon the value of the shape parameter Eq.~\eqref{Eq:GEVD1} can be categorized into three universality classes~\cite{Gumbel_1958}. For $\xi > 0$, Eq.~\eqref{Eq:GEVD1} reduces to the Fr\'echet class where the parent distribution decays as a power law. The case $\xi<0$ describes the  Weibull class for which the parent distribution decays faster than the power law. For these cases, the corresponding probability distribution function (PDF) is given by
\beqr
f(x, \mu, \beta, \xi) = & \dfrac{1}{\beta} \left( 1 + \xi \dfrac{x- \mu}{\beta}  \right)^{- (\xi + 1)/\xi} \nonumber \\
& \exp \Big\{ - \Big[ 1 + \xi \Big( \dfrac{x-\mu}{\beta} \Big) \Big]^{-1/\xi} \Big\} ~\label{Eq:PDF-GEV}
\eqnr

In the limit, $\xi \rightarrow 0$, Eq.~\eqref{Eq:GEVD1} takes the form

\beq
\mathcal{F}(x, \mu, \beta) = \exp \Big\{ -  \exp \Big(  \dfrac{x-  \mu}{\beta} \Big) \Big\}~\label{Eq:Gumbel-CDF}
\eqn

which describes the {\it Gumbel} class, and the corresponding PDF is given by
\beq
f(x, \mu, \beta) = \dfrac{1}{\beta} \exp\Bigg\{ -\exp\left( - \dfrac{x - \mu}{\beta} \right) - \left( \dfrac{x - \mu}{\beta} \right) \Bigg\}~\label{Eq:Gumbel-PDF}
\eqn

\section{Models}~\label{sec-model}

\subsection{Stochastic Manna Model}~\label{model-manna}

The SMM is a stochastic version of the sandpile model, introduced by S. S. Manna~\cite{Manna_1991}, and later, its modified version was proposed by D. Dhar~\cite{Dhar_1999}. The SMM is another well-studied model with robust scaling behavior. It is a paradigmatic example of a self-organized critical (SOC) system that belongs to the Manna universality class, distinct from the Abelian sandpile and directed percolation classes~\cite{Dhar_2006, Christensen_2005}. It consists of particles distributed on a $d$-dimensional lattice, where each site can hold multiple particles. A site becomes active when it reaches a threshold of two or more particles, at which point it topples by transferring two particles to randomly selected neighboring sites. This redistribution process introduces stochasticity in the evolution of avalanches, distinguishing it from deterministic sandpile models. The system evolves through such toppling events until all sites have fewer than two particles, leading to a self-organized steady state.

The dynamics of SMM can be described on a $d$-dimensional lattice having system size $N=L^d$. Each site is associated with randomly distributed grains, say $h_i$. A site is considered to be unstable if $h_i \geq h_c$, and it will topple according to the following relaxation rules
\beqr 
h_i &\longrightarrow h_i - h_c \nonumber \\
h_j &\longrightarrow h_j + 1, \nonumber 
\eqnr

Here, $j$ denotes independently chosen random sites from $2d$ nearest neighbors such that $j=1,...,d$. In this study, we consider a two-dimensional $(d=2)$ model with $h_c =2$.

\subsection{Bak-Tang-Wiesenfeld Model}~\label{model-btw}
Another model that we consider in this study is a two-dimensional Abelian sandpile model~\cite{BTW, Tang_1988, Bak_1996} of lattice size $N=L^2$. Each lattice site is associated with randomly distributed grains, say $h_i$. This dynamical variable could be any physical quantity such as grain density, stress, height, or energy \etc~ To drive the system, we update a randomly chosen site $i$ with $E_i \rightarrow E_i + 1$ at each time step. The lattice is considered unstable if $h_i \ge h_c$, where $h_c$ is the threshold of the dynamical variable. If the lattice is unstable at the $i^{th}$ site, it follows the relaxation rules given by
\beqr 
h_i &\longrightarrow h_i - h_c \nonumber \\
h_j &\longrightarrow h_j + 1, \nonumber 
\eqnr 
where $j$'s are the nearest neighbors (left, right, top and bottom) with $h_c=4$. The grains can leave the system at the boundaries, indicating that the system is dissipative. This rule is repetitively applied until all the sites become stable and the avalanche (or activity) stops. To ensure the relaxation of the time scales, new driving occurs only after an ongoing avalanche is over. 

In both models, the observables of interest are the avalanche size $(s)$ and area $(a)$. Therefore, we parameterize the avalanches by variables $x \in \{s, a \}$: namely, the avalanche size $(s)$ defined as the total number of topplings and the avalanche area $(a)$ that gives the total area affected by the avalanche.
The finite size scaling form for the probability of the observables $x$ follows the power law with a cutoff~\cite{Yadav_2022}
\beq 
P(x,x_c) = \begin{cases}
            Ax_c^{-\theta} x^{-\tau_x},  &   \text{\ \ for \ \ } x \ll x_c \\
            \text{rapid change},         &   \text{\ \ for \ \ } x \approx x_c      
            \end{cases}~\label{Eq:PDF1}
\eqn
where $\tau_x$ is the critical exponent, $\theta$ is the scaling exponent, and the upper cutoff is approximated by $x_c \sim N^{D_x}$ where $D_x$ is the cutoff exponent. 

\section{Results}\label{Sec:USF}

\begin{figure}
    \centering
    \subfloat{\includegraphics[scale=0.60]{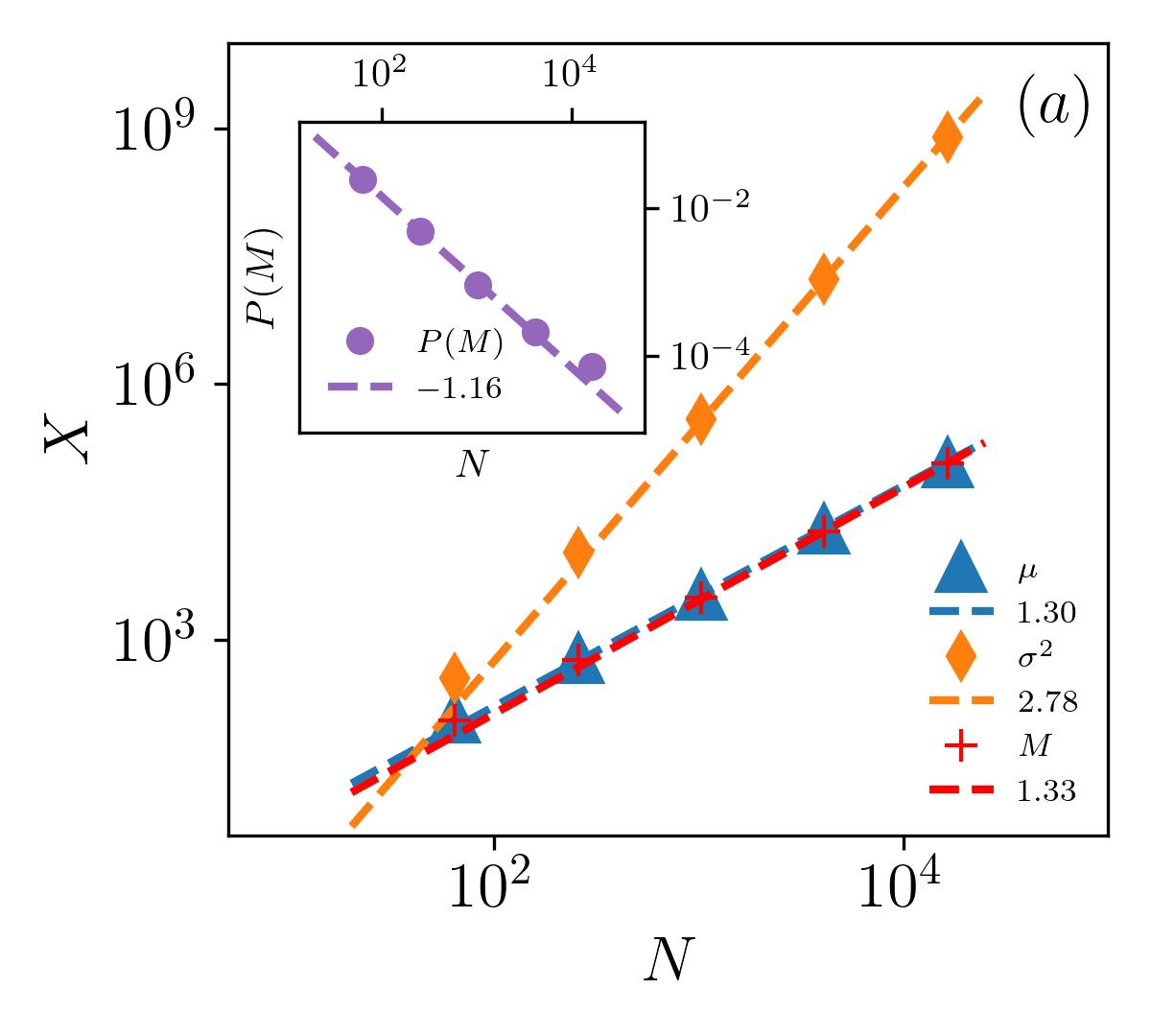}} \\ [-0.5ex]    \subfloat{\includegraphics[scale=0.60]{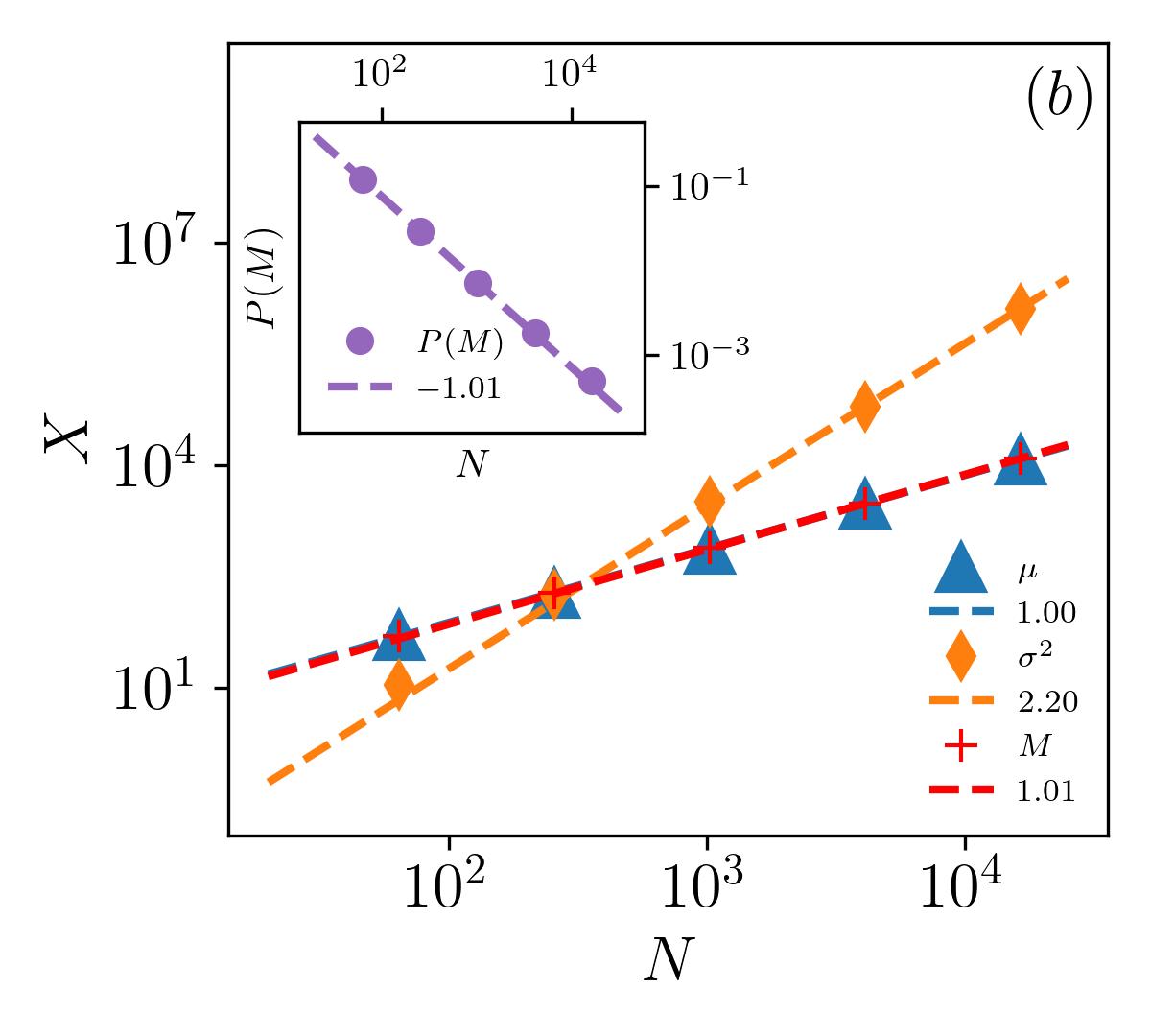}}
    \caption{In the SMM, the system scaling for mean, variance, mode and the probability for the mode (inset) of (a) extreme avalanche size (b) extreme avalanche area. The straight line represents the best-fit along with the fitted critical exponents.}~\label{Fig-Manna-X}
\end{figure}

\begin{figure}
    \centering
    \subfloat{\includegraphics[scale=0.8]{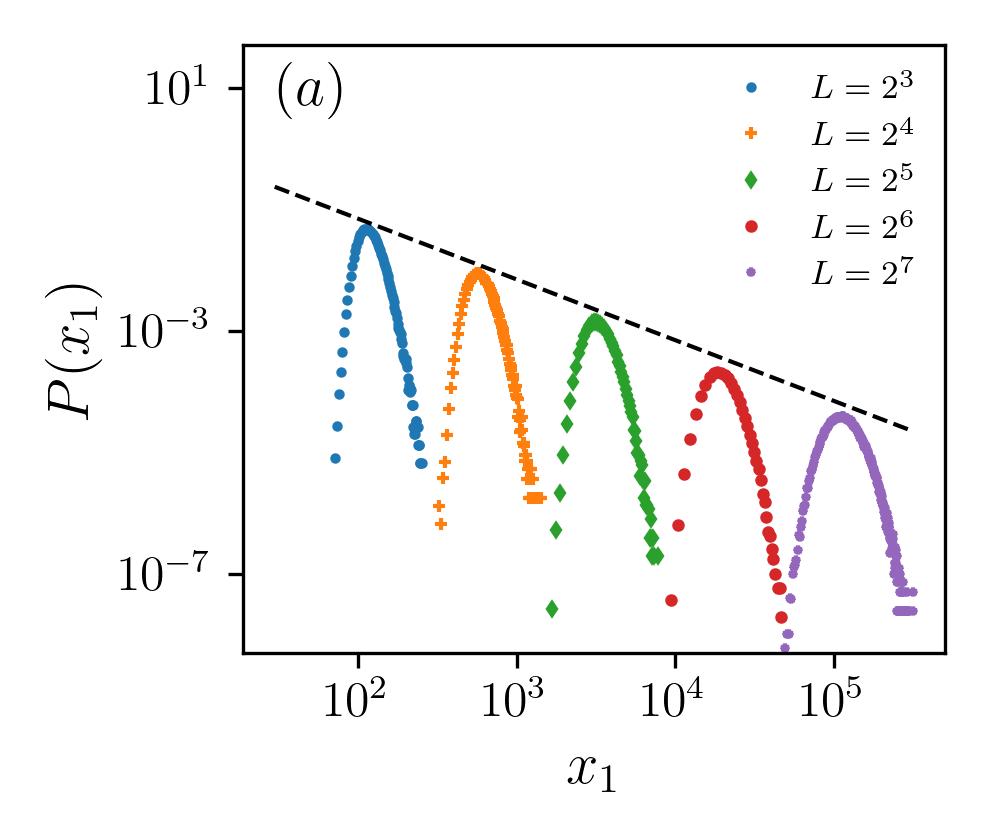}} \\ [-2ex]
    \subfloat{\includegraphics[scale=0.8]{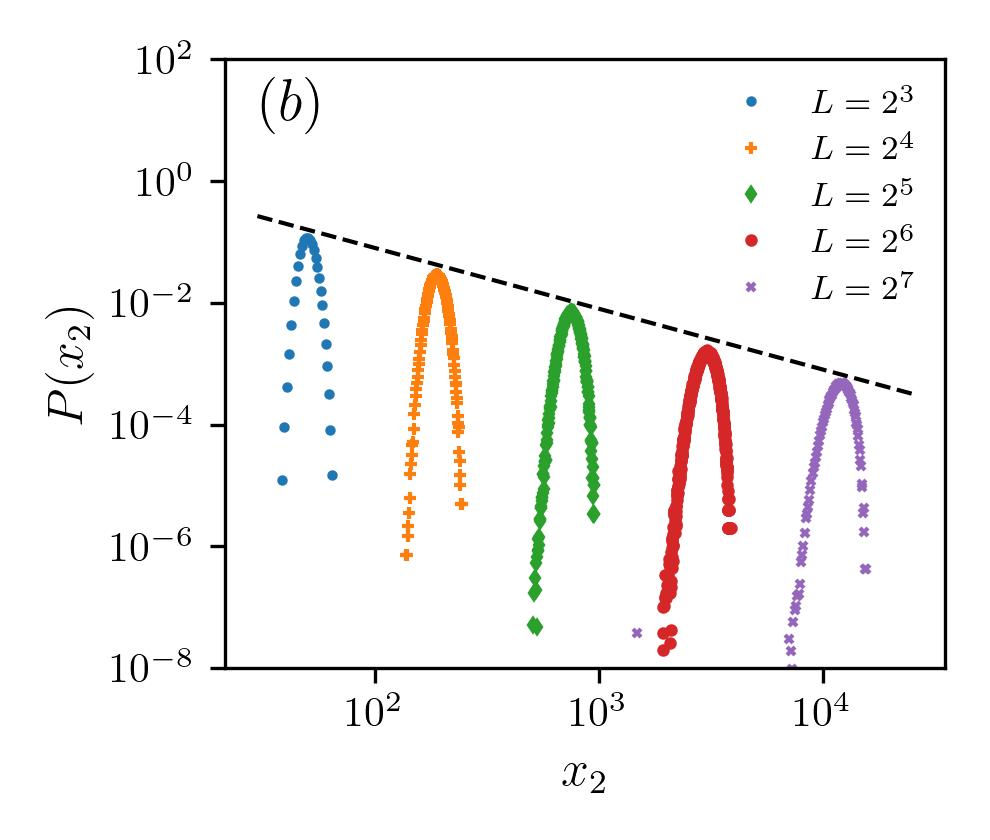}}
    \caption{The probability distribution $P(x)$ for (a) extreme avalanche size $x_1$ and (b) extreme avalanche area $x_2$ in case of the SMM for different system sizes $N=L^2$, where $L=2^3, 2^4,....,2^7$. The black dashed line has slope $-1$ which represents $P(M) \sim x^{-1}$.}~\label{Fig-Manna-Prob}
\end{figure}

In this section, we shall study the effect of system size on the statistics of extreme events. To generate these events, we use the block maxima technique. We consider a time series of the avalanche activity $z_i$, $i=1,\dots, T$, where $T$ is the length of the time series. The series is then divided into $M$ blocks ($M \in \mathbb{N}$) such that the avalanche activity is given by  $\{z_1, z_2,.....,z_M \}$. Extreme events are the maxima over the time series defined as $z^{max} = max\{z_1, z_2,.....,z_M\}$. The set $x_k = \{ z_i^{max},~\forall~i \}$ may follow a single peak extreme value distribution as discussed earlier (cf. Sec.~\ref{Sec:GEVT}). As shown in Fig~\ref{Fig-Manna-X}(a)-(b), the statistical quantities $X \in \{ \text{mean}~\mu,~\text{variance}~\sigma^2,~\text{and mode}~M \}$ for extreme activity show a system size scaling
\beqr
&\text{Mean~~~}& \mu = \dfrac{1}{T} \sum_{i=1}^T x_i \sim N^{\alpha_1}~\label{mean} \\
&\text{Variance~}& \sigma^2 = \la x^2 \ra - \la x \ra^2 \sim N^{2\alpha_2}~\label{var} 
\eqnr 
The probability distribution of extreme activities $P(M)$ [cf. Fig.~\ref{Fig-Manna-Prob}] shows a unimodal behavior with mode
\beq 
M \sim N^{\alpha_3}~\label{mode}
\eqn 
and the corresponding probability distribution of extreme activity decays with system size $N$ as [cf. Fig.~\ref{Fig-Manna-X}(inset)]
\beq 
P(M) \sim N^{-\alpha_4}~\label{peak}
\eqn

To understand the scaling behavior, we introduce a scaling variable
\beq 
u = \dfrac{x-M}{\sigma} = \dfrac{\Delta x}{\sigma}~\label{u}
\eqn 
Here, the scaling variable $u = \{u_1, u_2 \}$ corresponds to the extreme avalanche size $x_1$ and extreme avalanche area $x_2$ such that
$u_i = (x_i - M_{x_i})/\sigma_{x_i}$. This scaling variable rescales the peak of the probability distribution at $u=0$. Therefore, we expect the scaling function as
\beq 
F(u) = c \dfrac{P(x)}{P(M)}~\label{scal-fun}
\eqn 
A normalized probability distribution implies $\int P(x) dx=1$. Thus, plugging Eqs.~\eqref{var}-\eqref{u} into Eq.~\eqref{scal-fun} suggests
\beq 
\int P(x)dx \sim N^{\alpha_2 - \alpha_4} = 1 \nonumber
\eqn 
Thus, the normalization condition suggests $\alpha_2 = \alpha_4$. Similarly, the shifted extreme activity
\beq 
\la \Delta x \ra \sim N^{2\alpha_2 - \alpha_4} \sim N^{\alpha_1}
\eqn 
implies $2\alpha_2 - \alpha_4 = \alpha_1$ that yields $\alpha_1 = \alpha_2$. Since $u$ is independent of the system size $N$, thus $M \sim N^{\alpha_3 = \alpha_4}$~\cite{Chhimpa_2025}. Thus 
\beq 
\alpha_1 = \alpha_2 = \alpha_3 = \alpha_4 = \alpha
\eqn 
Then, from Eq.~\eqref{scal-fun} the probability distribution function behaves as
\beq 
P(x) \sim \dfrac{1}{N^\alpha} \mathcal{F} \left( \dfrac{x-M}{N^\alpha} \right) = \dfrac{1}{N^\alpha} \mathcal{F} \left( u \right) ~\label{Fig-Scale}
\eqn 

\begin{table}[h]
\caption{Critical exponents for extreme avalanche activities in case of the SMM and the BTW model.}~\label{Tab-CriExp1}
    \centering
    \renewcommand{\arraystretch}{1.3}
    \begin{tabular}{|p{2cm}<{\raggedright}|p{1.5cm}<{\raggedright}|c|c|c|c|}
        \hline
        {\bf Model} & {\bf Extreme activity} & $\alpha_1$ & $\alpha_3$ & $\alpha_3$ & $\alpha_4$ \\
        \hline
        \multirow{2}{*}{SMM} & Size & $1.30$ & $1.34$ & $1.33$ & $1.16$ \\
    
                                     & Area & $1.00$ & $1.10$ & $1.01$ & $1.01$ \\
        \hline
        \multirow{2}{*}{BTW} & Size & $1.19$ & $1.36$ & $1.13$ & $1.25$ \\
                                   & Area & $1.0$  & $1.10$ & $0.99$ & $1.10$\\
        \hline
    \end{tabular}
\end{table}

\begin{figure}
    \centering
    \includegraphics[scale=0.8]{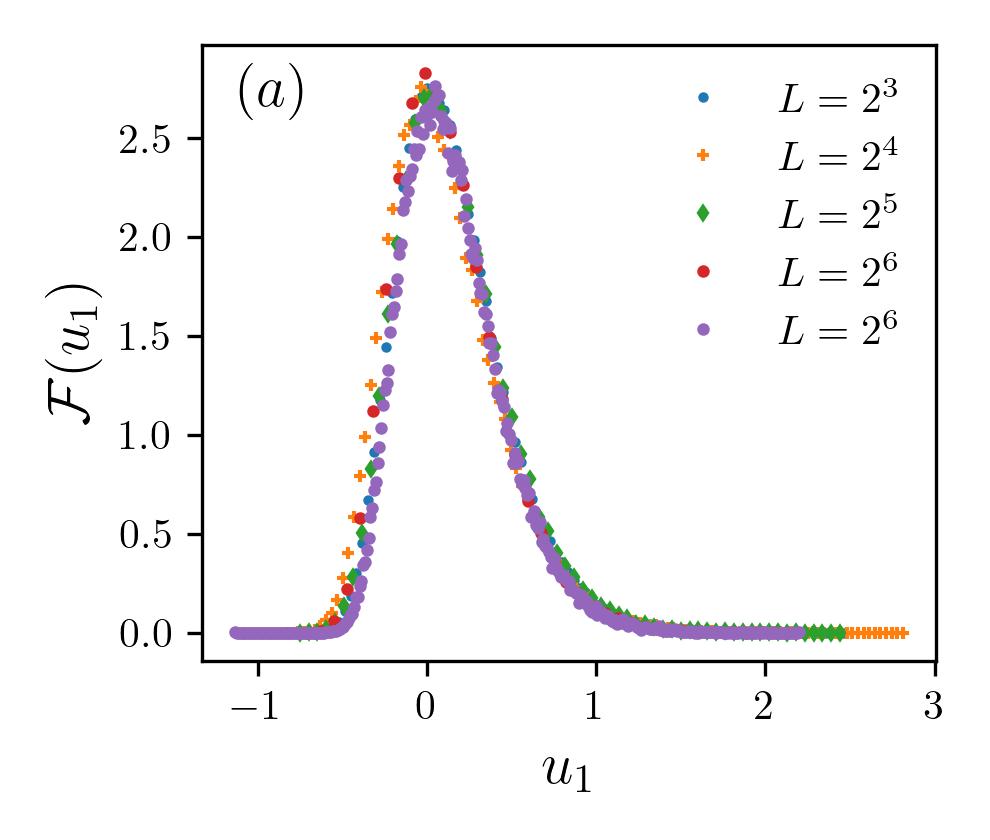} \\
    \includegraphics[scale=0.8]{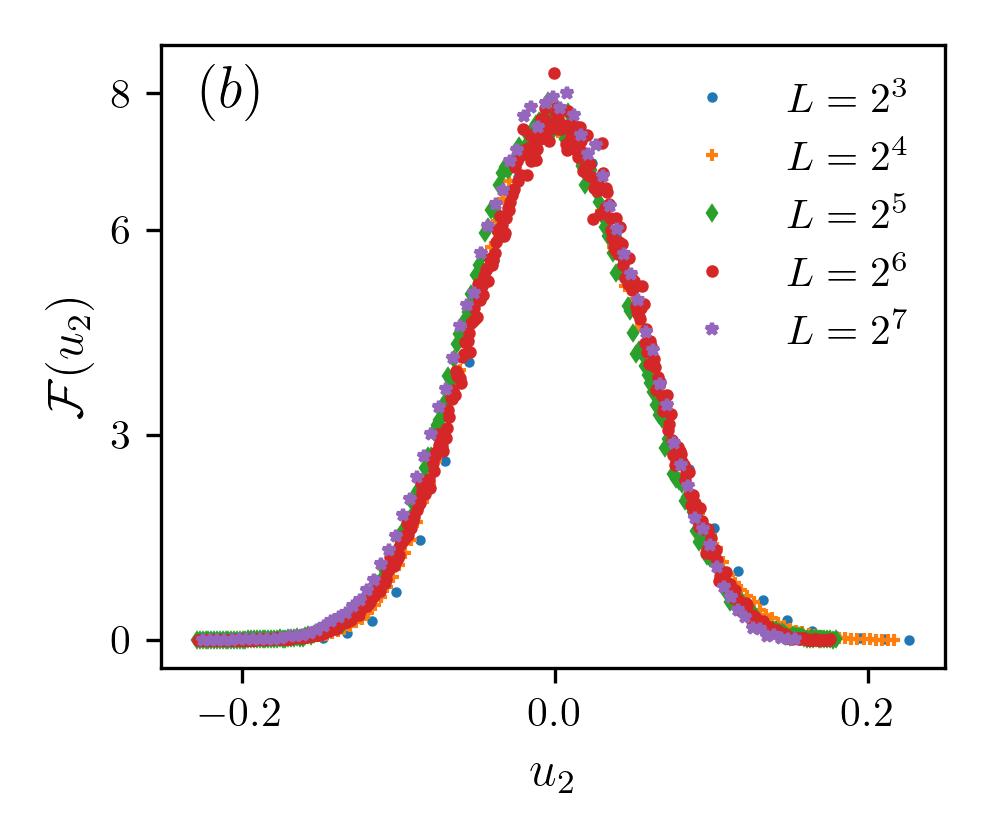}
    \caption{In SMM, the data collapse corresponding to Fig.~\ref{Fig-Manna-Prob} for the probability distribution of rescaled (a) extreme avalanche size $(u_1)$ and (b) extreme avalanche area $(u_2)$.}~\label{Fig-Manna-Col-F}
\end{figure}

Clearly, the plot between $N^\alpha P(x)$ and $u$ provides the scaling function $\mathcal{F}$ as shown in Figs.~\ref{Fig-Manna-Col-F} (a)-(b).

\begin{figure}
    \centering
    \includegraphics[scale=0.8]{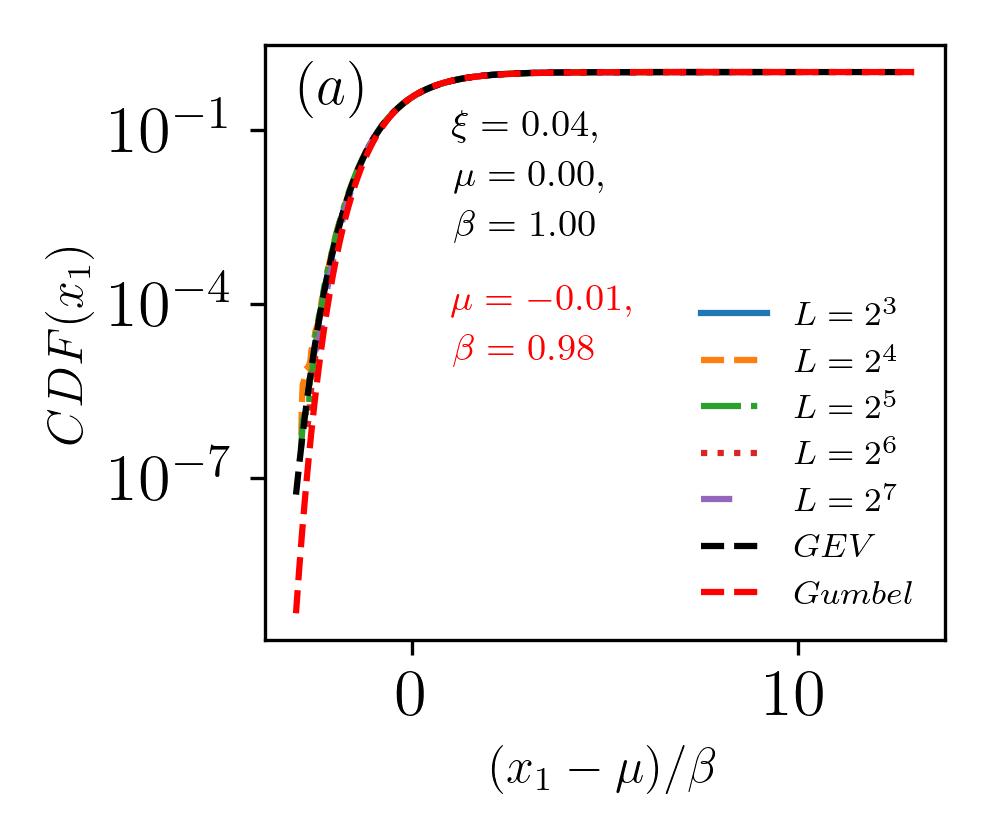} \\
    \includegraphics[scale=0.8]{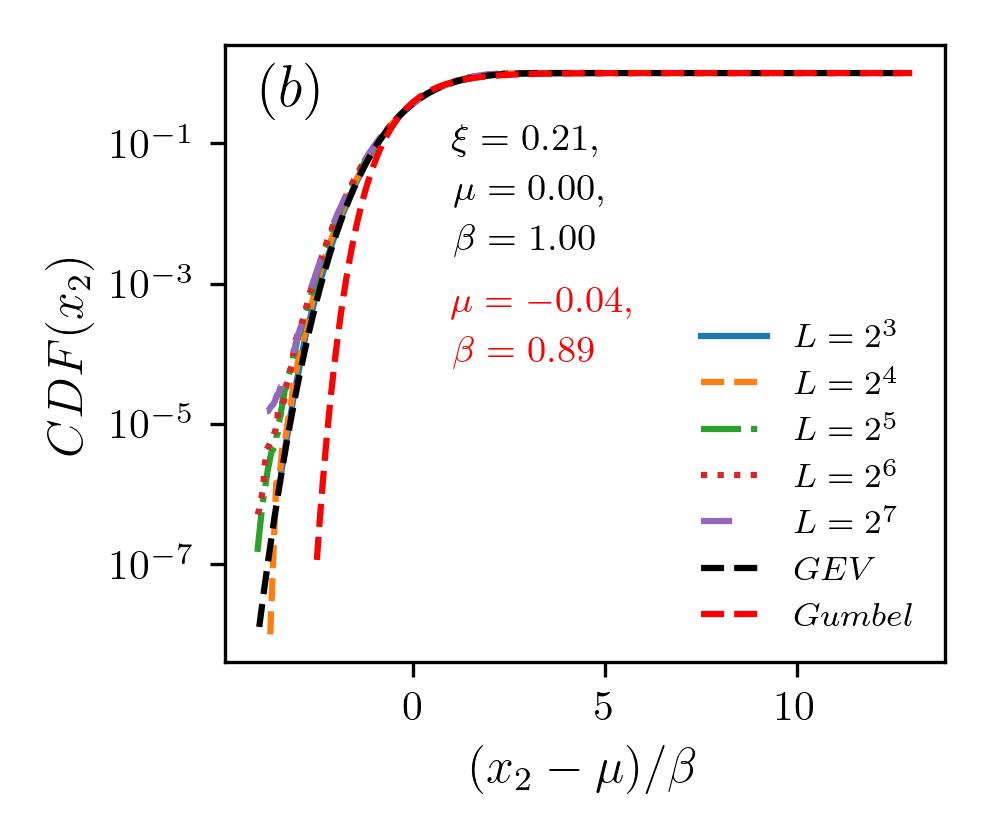}
    \caption{In SMM, the data collapse of the CDF corresponds to Fig.~\ref{Fig-Manna-Prob}. The red dashed line represents the fitted data with Gumbel family $(\xi=0)$ while the black dashed line represents the fitted GEV family with parameters $(\mu, \sigma, \xi)$. In both cases, the goodness of fit is $R^2>0.99$.}~\label{Fig-Manna-CDF}. 
\end{figure}

To understand the statistics more clearly, we study the cumulative probability distribution function $CDF(x) \sim \int P(x)dx$. Figs.~\ref{Fig-Manna-CDF}(a) and (b) show collapse for $CDF(x)$ for extreme avalanche size and area, respectively. We use the Levenberg-Marquardt algorithm (LMA)~\cite{LMA1, LMA2} to fit it with the Gumbel and GEV distribution function. We report the summary of the critical exponents in Table~\ref{Tab-CriExp1} and the fitting parameters of the scaling function in Table~\ref{Tab-GEVPar} for both SMM and BTW models. From the fitting parameters, we conclude that for both models, the extreme avalanche size follows the Gumbel family, while the extreme avalanche area shows a small positive shape parameter very close to the Gumbel family. 

\begin{table}[h]
\caption{The fitted parameters, describing the scaling function for the probability distribution of extreme avalanche activities for different SOC models. In all the cases, the goodness of fit is $R^2 > 0.99$.}~\label{Tab-GEVPar}
    \centering
    \renewcommand{\arraystretch}{1.2} 
    \begin{tabular}{|p{1.5cm}<{\raggedright}|p{1.7cm}<{\raggedright}|ccc|}
    \hline 
    &   &  \multicolumn{3}{c|}{GEV Parameters} \\
    {\bf Model} & {\bf Extreme activity} & $\xi$  & $\mu$ & $\beta$ \\ 
    \hline 
    \multirow{2}{*}{SMM} & Size & 0.04  & 0.00 & 1.00 \\
                                 & Area & 0.21  & 0.00 & 1.00 \\
    \hline 
    \multirow{2}{*}{BTW} & Size & 0.003 & -0.09 & 1.00 \\
                               & Area & 0.153 & -0.11 & 0.99 \\
    \hline 
    \end{tabular}
\end{table}

\section{Summary}~\label{Sec:Conclusion}
In summary, we study the extreme value statistics for the SMM and BTW models. With finite system size, the avalanche activities in the two models follow a power law with upper cut-off~\cite{Yadav_2022}. This upper cut-off (or FS) makes the study interesting. In the study, we examine various statistical characteristics of extreme avalanche activities. We apply statistical physics and EVT and record statistics over a range to explore the universal scaling of the extreme avalanche activities. For this, we use Monte-Carlo (MC) simulation to generate a set of realizations of extreme events for different system sizes. The statistical characteristics of extreme events, such as mean, variance, and the mode, follow the system size scaling characterized by the same critical exponent $(\alpha)$. We also present that the peak of the probability distribution for extreme avalanche activities shows a power law behavior with exponent $-1$. Employing the FS scaling with GEV theory, we suggest the scaling function (or data collapse) for the PDF and the CDF of extreme events. The data collapse fits with the GEV distribution having $\xi = 0$ for extreme avalanche sizes and $\xi>0$ for extreme avalanche area.

Thus, the present study is helpful in exploring the system size dependence of the EVD associated with the extreme events. This provides significant insight into one of the intriguing issues associated with extreme activities in SOC. Further, the proposed scaling method is such that the universal scaling function is found to be independent of the system size in the form of a rescaled variable. This rescaled variable could be explained by the RG theory of extreme events~\cite{Calvo_2012, Gyorgyi_2010, Gyorgyi_2008}. Our findings could be extended to understand the nature of the SOC systems and could help us design schemes for dynamics governed by the SOC phenomena in natural systems.  We further examined how extreme events evolve as the block size varies by analyzing their statistical properties. Our findings indicate that changing the observation range does not significantly alter the evolution of extreme events in the system. Our proposed scaling function has an intimate connection between extreme activities appearing on different length scales. Similar ideas may be implemented in weather modeling~\cite{Yao_2022}, integrable turbulence~\cite{Suret_2016, Kraych_2019}, to develop a surrogate model for characterizing extreme events. Whether or not such techniques will be applicable to other situations, namely the geophysical phenomena or financial markets, is still an open question.

\section{Acknowledgment}

AQ gratefully acknowledges support from the INSPIRE Fellowship (DST/INSPIRE Fellowship/IF180689), awarded by the Department of Science and Technology (DST), Government of India. The authors thank Dr Avinash Chand Yadav (BHU, Varanasi) for his helpful suggestions. 
Computational resources were provided by the PARAM Rudra Supercomputing Facility at the Inter-University Accelerator Centre (IUAC), New Delhi, under the National Supercomputing Mission (NSM), Government of India.

\end{document}